# A universal description of pseudorapidity distributions in both nucleus-nucleus and *p-p* collisions at now available energy regions

Z. J. Jiang[*], H. P. Deng and Y. Huang

*College of Science, University of Shanghai for Science and Technology, Shanghai 200093, China*

Investigations have shown that the collective motion not only appears in nucleus-nucleus but also in *p-p* collisions. The best tool for depicting such collective motion is relativistic hydrodynamics. In this paper, the collective motion is assumed obeying the hydro model which integrates the features of Landau and Hwa-Bjorken theory and is one of a very few analytically solvable models. The fluid is then supposed freezing out into charged particles from a time-like hypersurface with a fixed time of $t_{\rm FO}$. The researches of present paper show that this part of charged particles together with leading particles, which, by conventional definition, carry on the quantum numbers of colliding nucleons and take away the most part of incident energy, can give a proper universal description to the pseudorapidity distributions of charged particles measured in both nucleus-nucleus and *p-p* collisions at now available energy regions.

PACS number(s): 25.75.Ag, 25.75.Ld, 24.10.Nz

Ⅰ. Introduction

In recent years, especially with the operations of BNL-RHIC and after CERN-LHC, the natures of matter created in nucleus or particle collisions have been undergoing a wide experimental and theoretical research. One of the most important achievements arrived at from this research is that the quark-gluon plasma formed in collisions is in a strongly coupled state as fluid, instead of being in a conventionally believed state as weakly interacting partonic gas [1-5]. The investigations also have shown that this kind of strongly coupled quark-gluon plasma has not only been created in nucleus but also in particle, such as *p-p* collisions, and the motion of this partonic fluid is nearly as an ideal one with a very little viscosity [6-34].

The best approach for describing the spatiotemporal evolution of fluid-like partonic matter is the relativistic hydrodynamics, which was first put forward by L. D. Landau in his pioneering work in 1953 [35]. Owing to the high degree of nonlinearity and interconnection of hydro equations, Monte Carlo simulations is, as usual, widely employed to deal with them especially for 2 or 3-dimensional expansions or situations including viscosity. In

---

[*] E-mail: jzj265@163.com

Monte Carlo simulations, beside a powerful calculation system, there also needs a sophisticated skill for avoiding instabilities in solving partial differential hydro equations. Furthermore, since the results come from a man-made and non-transparent software package, the relations between them and physical law are not direct and clear. On the contrary, the analytical methods, concerning the most essential and important elements affecting the physical phenomena *via* ideal assumptions, provide us the most basic law underlying. In addition, the concise and explicit form of exact solution is unmatchable by Monte Carlo simulations. Hence, despite facing tremendous difficulties, the finding of analytical solution of relativistic hydro equations has been being our unswerving pursuit of the goal. It is an important field in high energy physics.

Up to now, the analytical solutions of hydrodynamics are only limited to 1+1 expansion for ideal fluid with simple equation of state. The first exact solution of 1+1 hydrodynamics was given by I. M. Khalatnikov about 61 years ago [36], which is for an accelerated system being assumed as a massless ideal fluid and initially at rest. The solution was presented in a complicated integral form and later was used by Landau in his hydro model study and obtained the rapidity distributions of charged particles [37], which are in generally consistent with the observations made at BNL-RHIC [38-40]. The second exact solution of 1+1 hydrodynamics is given by R. C. Hwa about 41 years ago [41]. This solution is for an accelerationless system with Lorentz invariant initial condition. The result got in this way is simple and explicit. From this solution, J. D. Bjorken was able to get a simple estimate for the initial energy density achieved in collisions from the final observables [42]. This makes the energy density be measurable in experiment. It is the first and by now the only formula being widely recognized as one for estimating the energy density of matter created in collisions. Hence, it receives much attention. This is the reason why Hwa's theory is usually named as Hwa-Bjorken hydro model. However, since the free parameter in the formula has not been well fixed, how to determine the mentioned energy density is still an open problem. Moreover, the invariant rapidity distributions obtained from this model are at variance with experimental observations. Theoretically, such distributions are only the limiting cases of $\sqrt{s_{NN}} \to \infty$.

Along with the discovery of collective motion of partonic matter created in BNL-RHIC and CERN-LHC, the investigations of relativistic hydrodynamics have entered into a very active period, becoming one of the most popular subjects. At the same time, the analytical solutions of hydro equations have got into a golden stage of rapid developments and achieved a number of good results [8-18]. For example, by generalizing the relation between ordinary rapidity and space-time one, Ref. [9] integrates Landau and Hwa-Bjorken two models into one, becoming a model called as unified hydrodynamics in this paper and presenting a set of exact solutions. By taking advantage of the traditional scheme of Khalatnikov potential, Ref. [10] solved analytically the hydro equations and gave a pack of simple exact solutions for an ideal fluid with linear equation of state. By taking into account the work done by the fluid elements on each other, Refs. [11-13] generalized the Hwa-Bjorken model for an accelerationless system to the model for an accelerated one, and obtained a class of exact analytical solutions of relativistic hydrodynamics.

One of the most important applications of 1+1 dimensional hydrodynamics is the analysis of the pseudorapidity distributions of charged particles in high physics. In the present paper, combing the effects of leading particles, we shall discuss such kinds of distributions in the framework of unified hydrodynamics for both nucleus-nucleus and *p-p* collisions at now available energy regions. The paper is organized as follows. First, in section II, we give a brief introduction to the theoretical model, presenting the exact solutions of unified hydro model and giving the rapidity distributions of charged particles resulted from both partonic fluid and leading particles. The model is then made a comparison in section III with experimental observations performed by PHOBOS Collaboration at BNL-RHIC in Au-Au collisions at $\sqrt{s_{NN}}$ =200 GeV [43] and by ALICE Collaboration at CERN-LHC in Pb-Pb collisions at $\sqrt{s_{NN}}$ =2.76 TeV [44]. In section IV, the theoretical model continues to undergo a test in *p-p* collisions at energy from 23.6 to 7000 GeV. The last section is traditionally about conclusions.

## II. A brief description of the model

The key ingredients of the model are as follows.

1. The evolution of fluid-like partonic matter created in collisions is dominated by the energy-momentum conservation

$$\frac{\partial T^{\mu\nu}}{\partial x^{\nu}} = 0, \quad \mu, \nu = 0, 1, \tag{1}$$

where $x^{\nu} = (x^0, x^1) = (t, z)$, $t$ is the time and $z$ is the longitudinal component of coordinates along beam direction. $T^{\mu\nu}$ is the energy-momentum tensor, which, for an ideal fluid, takes the form as

$$T^{\mu\nu} = (\varepsilon + p)u^{\mu}u^{\nu} - pg^{\mu\nu}, \tag{2}$$

where $g^{\mu\nu} = g_{\mu\nu} = \text{diag}(1,-1)$, $u^{\mu}$, $\varepsilon$ and $p$ are respectively the metric tensor, velocity, energy density and pressure of fluid. For a constant speed of sound, $\varepsilon$ and $p$ are related by the equation of state

$$\varepsilon = gp, \tag{3}$$

where $1/\sqrt{g} = c_s$ is the speed of sound. Eqs. (2) and (3) make Eq. (1) translate to

$$g\partial_+ \ln p = -\frac{(g+1)^2}{2}\partial_+ y - \frac{g^2-1}{2}e^{-2y}\partial_- y,$$
$$g\partial_- \ln p = \frac{(g+1)^2}{2}\partial_- y + \frac{g^2-1}{2}e^{2y}\partial_+ y, \tag{4}$$

where $\partial_+$ and $\partial_-$ are the compact notation of partial derivatives with respect to light-cone coordinates $z_\pm = t \pm z = x^0 \pm x^1 = \tau e^{\pm\eta_S}$, $\tau = \sqrt{z_+ z_-}$ is the proper time, $\eta_S = 1/2\ln(z_+/z_-)$ is the space-time rapidity and $y$ is the ordinary rapidity of fluid.

2. Eq. (2) is the complicated differential equations with high nonlinearity and coupling between variable $p$ and $y$. In order to solve it, the relation between $y$ and $\eta_S$ is generalized to the form [9]

$$2y = \ln u_+ - \ln u_- = \ln F_+(z_+) - \ln F_-(z_-), \tag{5}$$

where $u_\pm = e^{\pm y}$ are the light-cone components of 4-volicity, and $F_\pm(z_\pm)$ are the two arbitrary functions. In case of $F_\pm(z_\pm) = z_\pm$, Eq. (5) reduces to $y = \eta_S$, returning to the boost-invariant picture of Hwa-Bjorken. Otherwise, Eq. (5) describes the non-boost-invariant geometry of Landau. Accordingly, Eq. (5) unifies the Landau and Hwa-Bjorken hydrodynamics together. It plays a role of bridge between these two models.

By using Eq. (5), Eq. (4) becomes

$$g\partial_+ \ln p = -\frac{(g+1)^2}{4}\frac{f'_+}{f_+} + \frac{g^2-1}{4}\frac{f'_-}{f_+},$$
$$g\partial_- \ln p = -\frac{(g+1)^2}{4}\frac{f'_-}{f_-} + \frac{g^2-1}{4}\frac{f'_+}{f_-}, \tag{6}$$

where $f_\pm = F_\pm/H$, and $H$ is an arbitrary constant. After the above treatments, Eq. (6) becomes solvable and its solution is [9]

$$s(z_+, z_-) = s_0\left(\frac{p}{p_0}\right)^{\frac{g}{g+1}} = s_0 \exp\left[-\frac{g+1}{4}(l_+^2 + l_-^2) + \frac{g-1}{2}l_+ l_-\right], \tag{7}$$

where $s$ is the entropy density of fluid, and

$$l_\pm(z_\pm) = \sqrt{\ln f_\pm}, \quad y(z_+, z_-) = \frac{1}{2}(l_+^2 - l_-^2), \quad z_\pm = 2h\int_0^{l_\pm} e^{u^2} du, \tag{8}$$

$h = H/A$, and $A$ is an arbitrary constant fitting condition

$$F_\pm F''_\pm = \frac{A^2}{2}. \tag{9}$$

3. As the expansion of fluid lasts to the time of $t_{FO}$, the inelastic interactions between the particles in fluid cease, and the ratios of different kinds of particles maintain unchanged. At this moment, the collective movement of fluid comes to an end, and the fluid decouples or freezes out into the detected particles from a time-like hypersurface with a fixed time of $t_{FO}$. The rapidity distributions of entropy of fluid at this hypersurface take the form

$$\frac{dS}{dy} = su^\mu \frac{d\lambda_\mu}{dy}\bigg|_{t_{FO}} = su^\mu n_\mu \frac{d\lambda}{dy}\bigg|_{t_{FO}}, \tag{10}$$

where $n^\mu$ is the 4-dimensional unit vector normal to the hypersurface

$$n^\mu n_\mu = n_+ n_- = 1.$$

$d\lambda^\mu = d\lambda n^\mu$, and $d\lambda$ is the time-like infinitesimal length element along hypersurface

$$d\lambda = \sqrt{d\lambda^\mu d\lambda_\mu} = \sqrt{-dz^+ dz^-}, \tag{11}$$

where the minus sign accounts for the time-like characteristic of $d\lambda$. Considering that the number of charged particles is proportional to the amount of entropy, from solution (7) and Eq. (10) we can obtain the rapidity distributions of produced charged particles [9]

$$\frac{dN_{Fluid}}{dy} = Ce^{-(g-1)(l_+-l_-)^2/4} \frac{\partial_+\phi e^y + \partial_-\phi e^{-y}}{\partial_+\phi l_- e^y + \partial_-\phi l_+ e^{-y}}\bigg|_{t_{FO}}, \tag{12}$$

where $C$ is an overall normalization constant. $\phi$ stands for an arbitrary time-like hypersurface.

4. The right-hand side of Eq. (12) is evaluated on the time-like hypersurface with the time equaling $t_{FO}$. Such hypersurface can be therefore taken as

$$\phi(z_+, z_-) = t_{FO} = \frac{1}{2}(z_+ + z_-) = C, \tag{13}$$

where $C$ is an arbitrary constant. This equation gives

$$\partial_\pm \phi = \frac{1}{2}. \tag{14}$$

Thus, Eq. (12) turns into

$$\frac{dN_{Fluid}}{dy} = C\frac{e^{-(g-1)(l_+-l_-)^2/4}}{l_- + l_+ + (l_- - l_+)\tanh y}. \tag{15}$$

5. In nucleus or particle collisions, apart from the freeze-out of fluid, leading particles also have certain contribution to the measured charged particles. Leading particles are believed to be formed outside the nucleus, that is, outside the colliding region [45, 46]. The motion and generation of leading particles are therefore free from hydro descriptions. As we have argued before that the rapidity distribution of leading particles takes the Gaussian form [15, 16]

$$\frac{dN_{Lead}}{dy} = \frac{N_{Lead}}{\sqrt{2\pi}\sigma}\exp\left\{-\frac{[|y|-y_0]^2}{2\sigma^2}\right\}, \qquad (16)$$

where $N_{Lead}$, $y_0$ and $\sigma$ are respectively the number of leading particles, central position and width of distribution. This conclusion comes from the consideration that, for a given incident energy, different leading particles resulting from each time of collisions have approximately the same amount of energy or rapidity. Then, the central limit theorem [47, 48] guarantees the plausibility of above argument. Actually, experimental measurements have shown that any kind of charged particles presents a good Gaussian rapidity distribution [38-40].

$y_0$ in Eq. (16) is the average position of leading particles. It should increase with incident energies and centrality cuts. The value of $\sigma$ relies on the relative energy or rapidity differences among leading particles. It should not, at least not apparently depend on the incident energies, centrality cuts and even colliding systems. The specific values of $y_0$ and $\sigma$ can be determined by tuning the theoretical predictions to experimental data.

By definition, leading particles mean the particles inheriting the quantum numbers of colliding nucleons. Hence, the number of leading particles is equal to that of participants. For *p-p* collisions, there are only two leading particles. They are separately in projectile and target fragmentation region. For an identical nucleus-nucleus collision, the number of leading particles

$$N_{Lead} = \frac{N_{Part}}{2}, \qquad (17)$$

where $N_{Part}$ is the total number of participants, which can be determined in theory by Glauber model [49, 50].

## III. Comparison with experimental measurements in nucleus-nucleus collisions

From Eqs. (15) and (16), we can get the pseudorapidity distributions of charged particles in nucleus-nucleus collisions. The results are shown in Figs. 1 and 2, which are for different centrality Au-Au and Pb-Pb collisions at $\sqrt{s_{NN}}$ =200 GeV and 2.76 TeV, respectively. The solid dots in the figures are the experimental measurements [43, 44]. The dashed curves are the results got from unified hydrodynamics of Eq. (15). The dashed-dotted curves are the results obtained from leading particles of Eq. (16). The solid curves are the sums of dashed and dashed-dotted curves. It can be seen that the combined contributions from both unified hydrodynamics and leading particles match up well with experimental data.

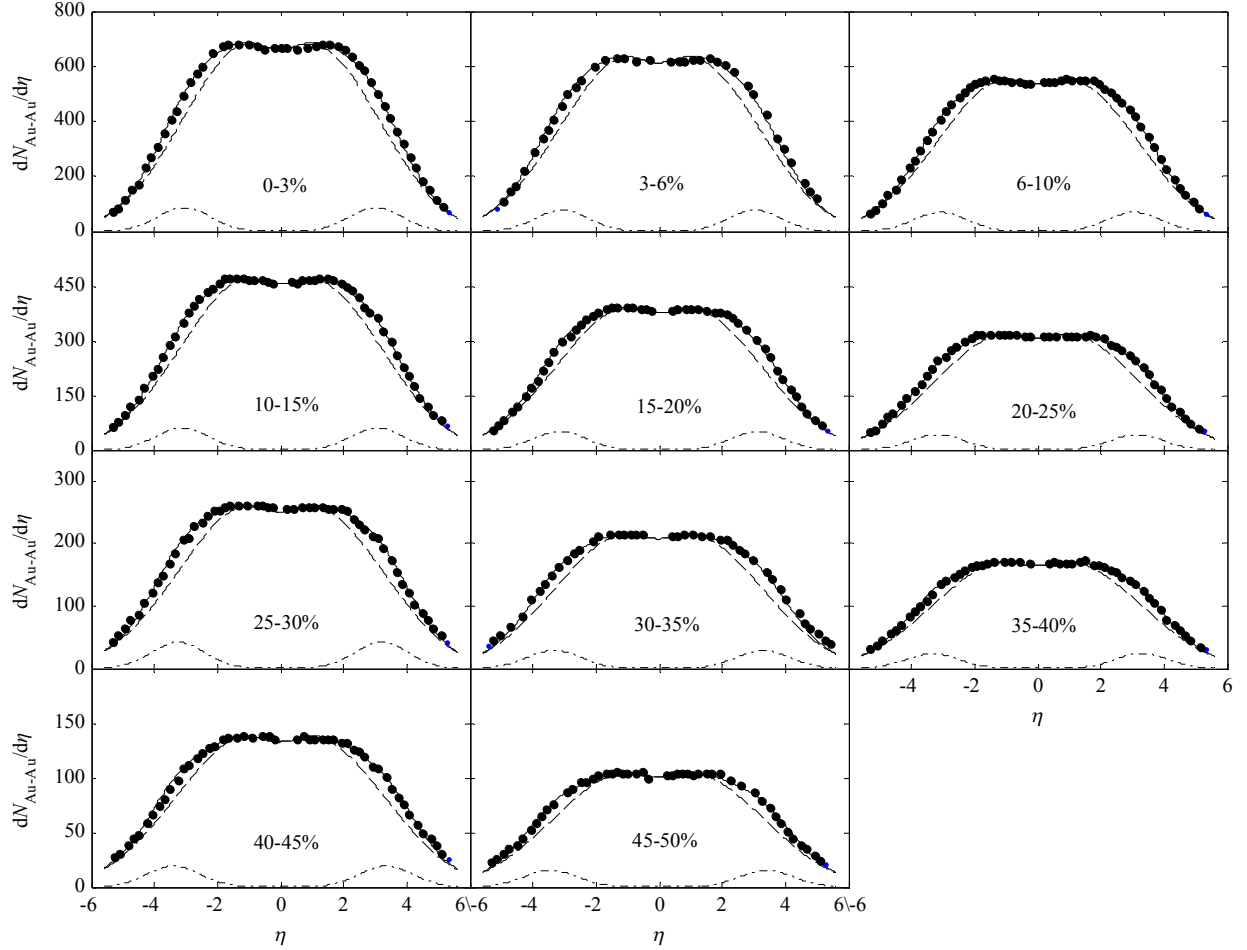

FIGURE 1: The pseudorapidity distributions of produced charged particles in different centrality Au-Au collisions at $\sqrt{s_{NN}} = 200$ GeV. The solid dots are the experimental measurements [43]. The dashed curves are the results from unified hydrodynamics of Eq. (15). The dashed-dotted curves are the results from leading particles of Eq. (16). The solid curves are the sums of dashed and dashed-dotted curves.

In calculations, the factor $g$ in Eq. (3) takes separately the values of 6.6 and 4.5 for different centrality Au-Au and Pb-Pb collisions. That is, it decreases with increasing energies. It should change like this. Since $g$ decreases with increasing temperature of partonic fluid [51], and the larger the energy, the higher is the temperature of the created fluid. The parameter $h$ in Eq. (8) takes the values of $5.4 \times 10^{-4}$ - $0.2 \times 10^{-4}$ and $1.0 \times 10^{-5}$ - $0.3 \times 10^{-5}$ for centrality cuts from small to large in Au-Au and Pb-Pb collisions, respectively. It can be seen that $h$ decreases with increasing energies and centrality cuts. It should vary in this way. Since the larger the energy and centrality cut, the more transparent the nucleus becomes. The produced charged particles will then be located in a wider rapidity region. While, the region of rapidity distributions of Eq. (15) is mainly determined by $h$. The wider the region of rapidity distributions, the smaller is the value of $h$. This can be seen in Fig. 3, which is for Au-Au collisions normalized to 0-3% centrality cut. The wider curve is the corresponding rapidity distribution in Fig. 1, obtained by using $h = 5.4 \times 10^{-4}$. The narrower one is for $h = 5.4 \times 10^{-2}$. Hence, the region of rapidity distributions decreases with increasing $h$. This forms the basis for the above fitted $h$ values.

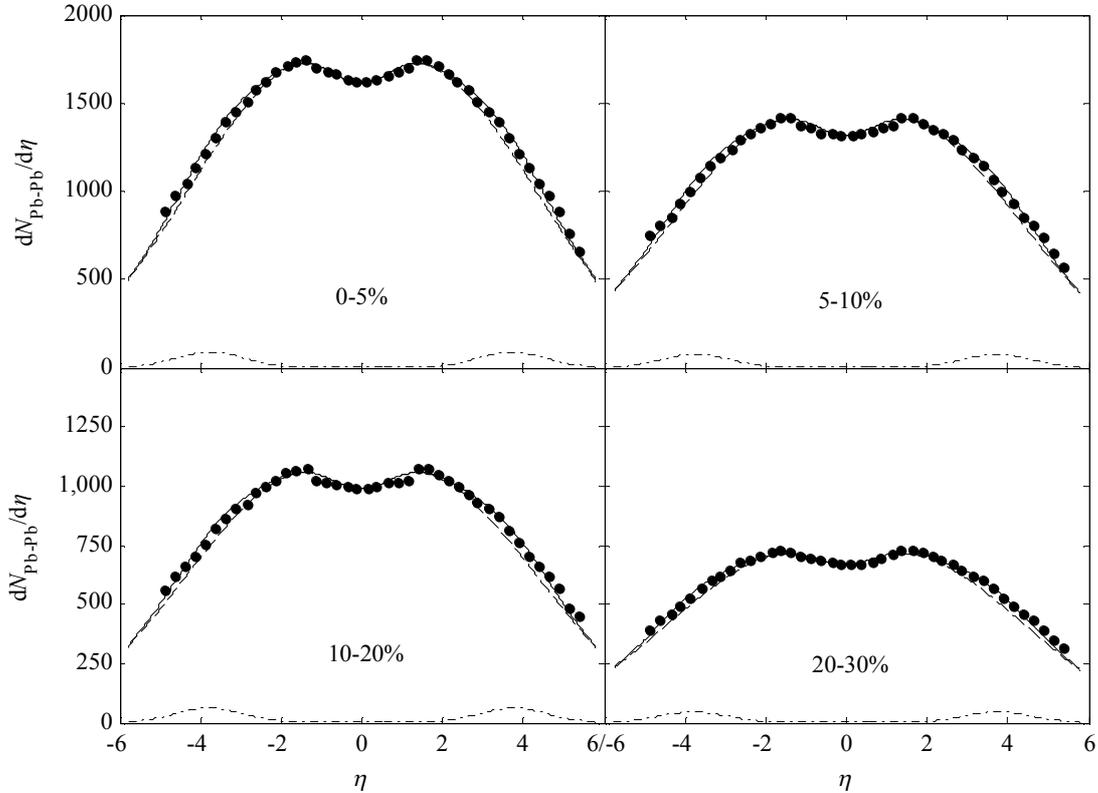

FIGURE 2: The pseudorapidity distributions of produced charged particles in different centrality Pb-Pb collisions at $\sqrt{s_{NN}} = 2.76$ TeV. The solid dots are the experimental measurements [44]. The dashed curves are the results from unified hydrodynamics of Eq. (15). The dashed-dotted curves are the results from leading particles of Eq. (16). The solid curves are the sums of dashed and dashed-dotted curves.

The central parameter $y_0$ in Eq. (16) takes the values of 2.91-3.30 and 3.61-3.65 for centrality cuts from small to large in Au-Au and Pb-Pb collisions, respectively. As mentioned above, it increases with centrality cuts and energies. The width parameter $\sigma$ in Eq. (16) takes a constant of $\sigma = 0.85$ for different centrality cuts in both Au-Au and Pb-Pb collisions. As stated above, it is irrelevant to centrality cuts, incident energies and colliding systems.

## IV. Comparison with experimental measurements in *p-p* collisions

Compared with nucleus-nucleus collisions, *p-p* collisions are a relatively simpler and smaller system, and therefore produce the matter in a much tinier volume. If such matter expands in the same way as that generated in nucleus-nucleus collisions has been experiencing endless disputations. Such situation has changed since a series of findings in recent years [19-34]. Refs. [19-23] corroborated the formation of quark-gluon plasma in *p-p* collisions. The measurements from CMS Collaboration at CERN-LHC have shown that [24], just like in nucleus-nucleus collisions, the ridge structures, the signal of collective motion, also appear in *p-p* collisions. Refs. [25, 26] further substantiated that these observed ridge structures can be well understood in the framework of hydrodynamics. In Ref. [27], the STAR Collaboration at BNL-RHIC has measured the HBT (Hanbury-Brown-Twiss) radii for *p-p* collisions as a function of multiplicity. The measured results are well favored by the predictions of hydro models made in Res. [28-30]. In Refs. [31, 32], E. K. G. Sarkisyan *et al.* utilized the combined model of Landau hydrodynamics plus combinations of the constituent quarks in participants to deal with the multihadron productions and transverse momentum distributions in *p-p* ($\bar{p}$) collisions. By employing the hydro solutions known as Gubser flow [18], Ref. [33] demonstrated the collective radial flow of high multiplicity *p-p* collisions. In our previous work [34], we used the evolution-dominated hydrodynamics together with the effects of leading particles to describe the pseudorapidity distributions of charged particles produced in *p-p* collisions. At present paper, from Eqs. (15) and (16), we can also get such distributions. The results are shown in Fig. 4 for *p-p* collisions at energies from 23.6 to 900 GeV. The solid dots are the experimental measurements [52-54]. The dashed curves are the results from unified hydrodynamics of Eq. (15). The dashed-dotted curves are the results from leading particles of Eq. (16). The solid curves are the sums of dashed and dashed-dotted curves. It can be seen that the theoretical results are in well consistent with experimental data.

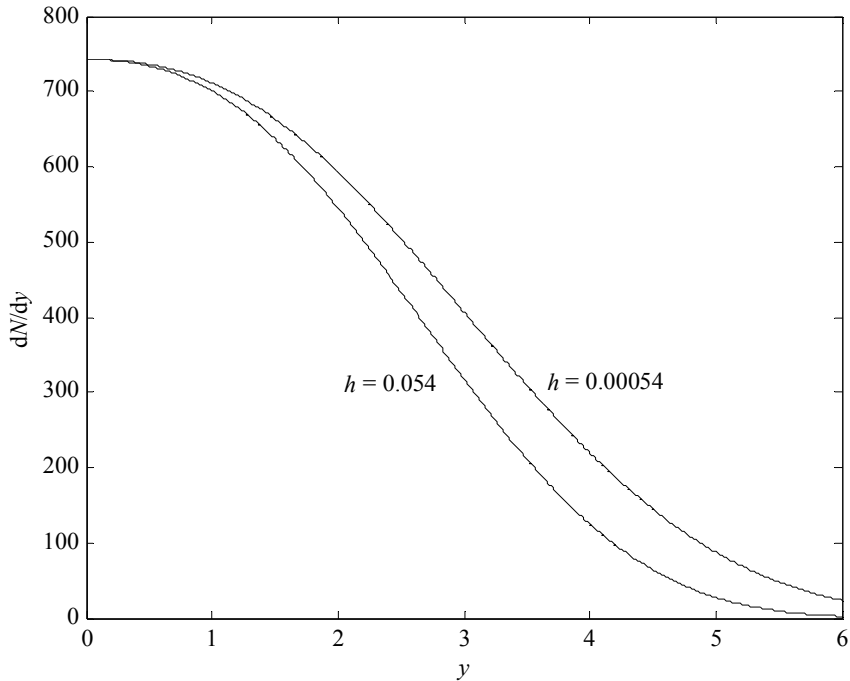

FIGURE 3: The rapidity distributions of produced charged particles for two different *h* values in Au-Au collisions normalized to 0-3% centrality cut.

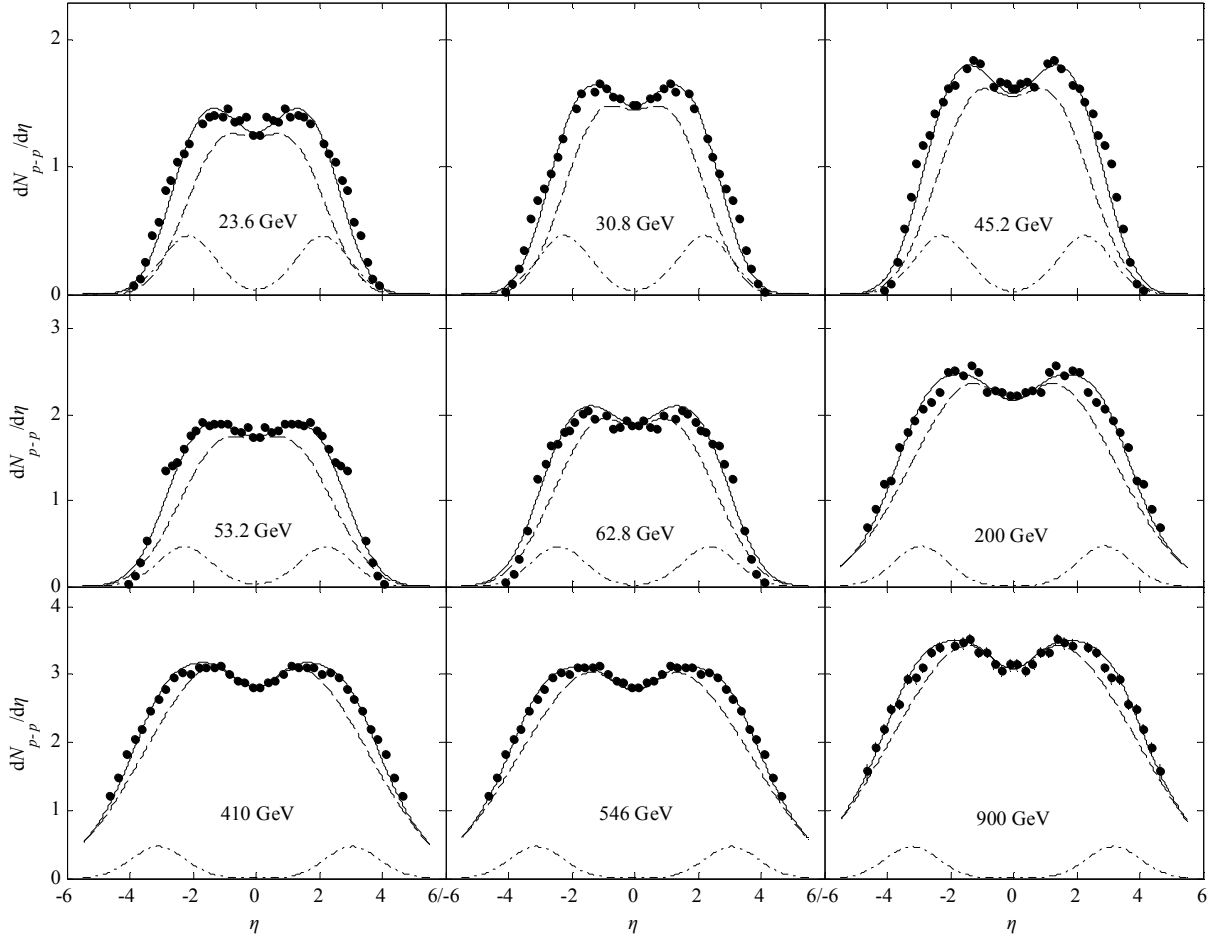

FIGURE 4: The pseudorapidity distributions of produced charged particles in *p-p* collisions at energies from $\sqrt{s} = 23.6$ to 900 GeV. The solid dots are the experimental measurements [52-54]. The dashed curves are the results from unified hydrodynamics of Eq. (15). The dashed-dotted curves are the results from leading particles of Eq. (16). The solid curves are the sums of dashed and dashed-dotted curves.

In calculations, the width parameter $\sigma$ in Eq. (16) takes the same value as that in nucleus-nucleus collisions, that is, $\sigma = 0.85$. This certifies the above argument again that $\sigma$ is independent of energies and colliding objects. The factor $g$, $h$ and $y_0$ take the values of 8.4-5.2, 18.4- $4.0 \times 10^{-5}$ and 1.91-3.00, respectively, for energies from low to high. Just as in nucleus-nucleus collisions, $g$ and $h$ decrease with increasing energies, and $y_0$, on the contrary, increases with energies. Their variations against energies are presented in Fig. 5. The solid dots represent the above fitted values. The solid curves are respectively drawn from relations

$$\begin{aligned}
g &= 11.3311 - 0.5324 \ln s + 0.0065 \ln^2 s, \\
\ln h &= 23.3049 - 3.7056 \ln s + 0.0877 \ln^2 s, \\
y_0 &= 0.3657 + 0.2709 \ln s - 0.0055 \ln^2 s,
\end{aligned} \quad (18)$$

where $\sqrt{s}$ is in the unit of GeV. It can be seen that all the solid dots are well seated on the curve. The stars and circles in this figure are the corresponding predictions for *p-p* collisions at CERN-LHC energies of 2.36 and 7 TeV, respectively. From these predictions, we can get the pseudorapidity distributions of produced charged particles in these two situations, and the results are shown in Fig. 6. The solid dots are the experimental data [55]. The dashed curves are the predicted results from unified hydrodynamics of Eq. (15). The dashed-dotted curves are the predicted results from leading particles of Eq. (16). The solid curves are the sums of dashed and dashed-dotted curves. It can be seen that the theoretical predictions agree well with the available measurements in the mid-pseudorapidity regions.

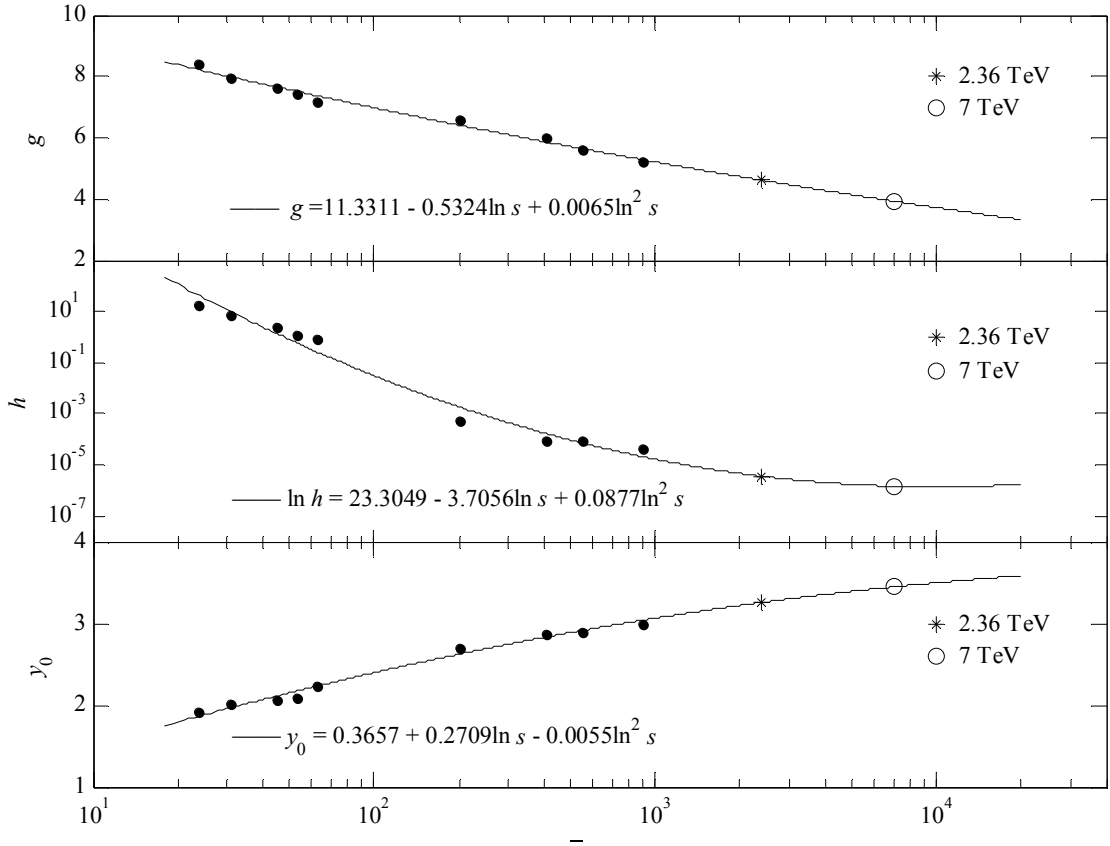

FIGURE 5: The variations of $g$, $h$ and $y_0$ against $\sqrt{s}$. The solid dots represent the fitted values given in the text. The stars and circles are the predictions for *p-p* collisions at CERN-LHC energies of $\sqrt{s}$ = 2.36 and 7 TeV, respectively.

## V. Conclusions

The charged particles in nucleus and particle collisions are supposed having the same producing mechanism. That is, one part is from the freeze-out of fluid-like partonic matter, the other is from leading particles.

The fluid-like partonic matter is assumed evolving according to the relativistic hydrodynamics, which incorporates the characteristics of Landau and Hwa-Bjorken hydro model *via* a generalizing relation between ordinary rapidity $y$ and space-time rapidity $\eta_S$. This is one of a very few hydro models which can be solved analytically. The solutions can then be utilized to formulate the rapidity distributions of charged particles frozen out from a time-like hypersurface with a fixed time of $t_{FO}$. In the derived formula, there are two parameters $g$ and $h$. Known from comparing with experimental measurements, both $g$ and $h$ decrease with increasing energies, and in nucleus-nucleus collisions, $g$ maintains unchanged for different centrality cuts, while, $h$ decreases with increasing centrality cuts.

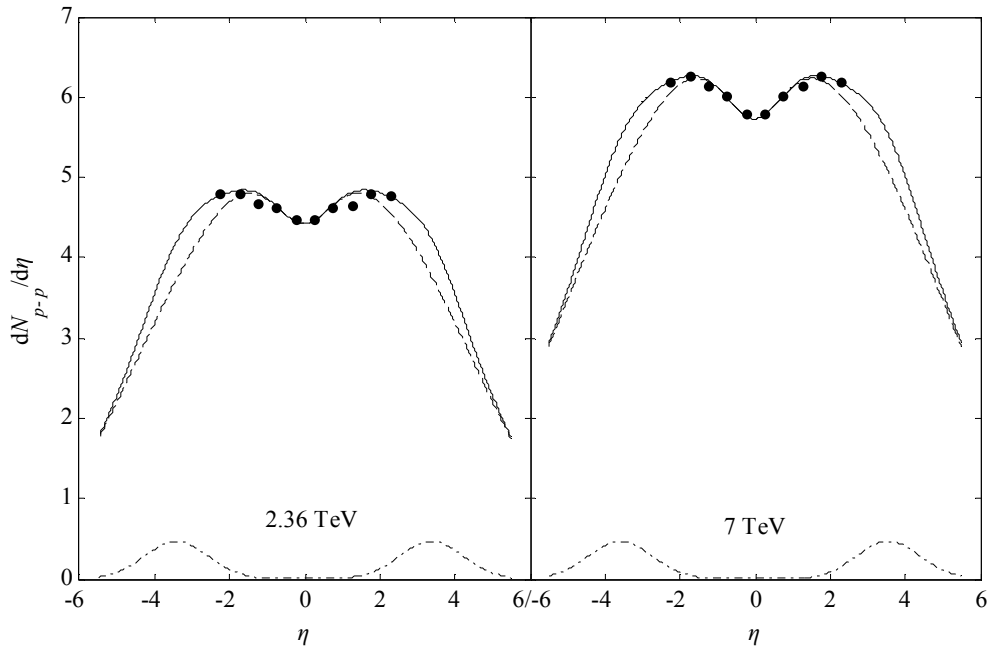

FIGURE 6: The pseudorapidity distributions of produced charged particles in *p-p* collisions at energies of $\sqrt{s} = 2.36$ and 7 TeV. The solid dots are the experimental measurements [55]. The dashed curves are the predicted results from unified hydrodynamics of Eq. (15). The dashed-dotted curves are the predicted results from leading particles of Eq. (16). The solid curves are the sums of dashed and dashed-dotted curves.

For leading particles, the rapidity distributions of them are, as usual, believed possessing the Gaussian form normalized to a constant equaling the number of participants, which can be figured out in theory. This assumption is based on the consideration that, for a given incident energy, the leading particles have about the same energy, and coincides with the fact that any kind of the charged particles produced in collisions takes on well the Gaussian form of rapidity distribution. It is interested to note that the width of Gaussian rapidity distribution $\sigma$ is irrelevant to the energies, centrality cuts, and colliding systems.

Comparing with experimental measurements, we can see that the pseudorapidity distributions of produced charged particles in both nucleus-nucleus and *p-p* collisions at now available energy regions can be well described in a universal manner of unified hydrodynamics plus leading particles.

**Conflict of Interests**

The authors declare that there is no conflict of interests regarding the publication of this paper.

**Acknowledgments**

This work is partly supported by the Hujiang Foundation of China with Grant No. B14004; the Cultivating Subject of National Project with Grant No. 15HJPY-MS04 and Shanghai Key Lab of Modern Optical System.